\documentclass[pre,superscriptaddress,showpacs,twocolumn]{revtex4}
\usepackage{color}
\usepackage[usenames,dvipsnames]{xcolor}

\usepackage{graphicx}
\usepackage{color}
\usepackage{epsfig}
\usepackage{dcolumn}
\usepackage{bm}
\usepackage{mathrsfs}
\usepackage{multirow}
\usepackage[all]{xy}
\usepackage{pbox}
\usepackage{amssymb}

\def\(({\left(}
\def\)){\right)}
\def\[[{\left[}
\def\]]{\right]}

\newcommand{\Ss}{\mathcal S}

\newcommand{\Pp}{\mathcal P}

\newcommand{\Bx}{{\bf x}}
\newcommand{\Yy}{{\mathcal Y}}
\newcommand{\Ww}{{\mathcal W}}
\newcommand{\rref}{{\rho_{\tt ref}}}
\newcommand{\phiref}{{\phi_{\tt ref}}}
\newcommand{\rss}{{\rho_{\tt ss}}}

\newcommand{\Tx}{\mathcal T}

\newcommand{\beq}{\begin{equation}}
\newcommand{\eeq}{\end{equation}}
\newcommand{\vn}{\mathbf{n}}
\newcommand{\la}{\langle}
\newcommand{\ra}{\rangle}
\newcommand{\vx}{{\bf{x}}}

\begin{document}  

\title{Infinite family of second-law-like inequalities} 

\author{Carlos P\'erez-Espigares}
\email[]{cpespigares@onsager.ugr.es}
\affiliation{Departamento de Electromagnetismo y F\'{\i}sica 
de la Materia,
Universidad de Granada, Granada 18071, Spain}

\author{Alejandro B. Kolton}
\email[]{koltona@cab.cnea.gov.ar}
\affiliation{CONICET, Centro At\'omico Bariloche, 8400 San Carlos de Bariloche, R\'io Negro, Argentina}

\author{Jorge Kurchan}
\email[]{jorge@pmmh.espci.fr}
\affiliation{PMMH-ESPCI, CNRS UMR 7636, 10 rue Vauquelin,
 75005 Paris, France}

\date{\today}

\begin{abstract}
The probability distribution function for an out of equilibrium system may sometimes be approximated
by a physically motivated ``trial'' distribution. 
A particularly interesting case is when a driven  system (e.g., active matter) is approximated by a thermodynamic one.
We show here that every set of trial distributions yields an inequality 
playing the role of a generalization of the second law.
The better the approximation is, the more constraining the inequality becomes: this 
suggests a criterion for its accuracy, as well as an optimization procedure 
that  may be implemented numerically and even experimentally. The 
fluctuation relation behind this inequality, -a natural and practical 
extension of the Hatano-Sasa theorem-, does not rely on the 
a priori knowledge of the stationary probability distribution. 
\end{abstract}


\maketitle 

\section{Introduction}
\label{sec:introduction}
A recurring strategy applied to  out of equilibrium systems is to represent the complex 
energy and dissipation sources by a bath with good equilibrium thermal properties.
Two examples are the Edwards ``thermodynamic`` approach to granular matter \cite{edw}  and recent developments
for active matter
(see Refs. \cite{jor,leticia,wolynes} for recent examples), in which the combination of rapid energy bursts 
and friction is mimicked by a thermal bath. The aim of such pursuits is not necessarily to make the 
problem more easily solvable,
but rather  to cast it in a form that provides  thermodynamic intuition and constraints.
In this paper we derive some  simple relations that help  make this mapping more systematic and controlled.
The method is based on the use of  inequalities of the form of the  second law, associated with each guess for the  distribution function.

In these last two decades there has been a  development of a family of relations valid for out of equilibrium systems~\cite{reviews},
including the  fluctuation theorem in its various forms~\cite{EvCoMo,GaCo,Ku,Hurtado} and the Jarzynski~\cite{Ja} and 
Crooks~\cite{Cr} relations. 
A later and extremely simple result is the Hatano-Sasa equality~\cite{HaSa}, 
which applies to systems that  are continuously driven by time-dependent potentials inducing currents, 
  so that even  the stationary regimes are out of equilibrium. Their result is essentially a version of Jarzynski equality and the second principle,
but   with the  energy replaced by the logarithm of the stationary distribution.

In this paper we derive an extension of the Hatano-Sasa 
 theorem for Markovian systems, which 
has the practical advantage that it {\it does not rely on the a priori knowledge of the stationary 
probability distribution}. 
Instead, arbitrary ``trial'' smooth 
distributions can be used, thus  allowing one to treat  
systems whose stationary distribution is either (i) 
too difficult to calculate, as generically 
occurs in out of equilibrium systems with many 
degrees of freedom, or (ii) unwieldy, 
as, for instance, in the deterministic limit, 
where the nonequilibrium steady-state distributions are nonzero only over a 
fractal support.
Our approach leads in particular 
to an inequality
 that 
can be used as a variational principle for improving, 
in a
 controlled way, physically motivated approximations to
nonequilibrium steady-state distributions. 
The optimization  procedure might 
be implemented numerically or even experimentally.
As an illustration, we numerically 
approximate the 
stationary distribution of a paradigmatic
non-equilibrium driven system with many degrees 
of freedom, the simple symmetric exclusion process in one dimension.  

Just as in the case of the Hatano-Sasa equality~\cite{HaSa, Esposito},
there is a   ``dual'' (or adjoint)   ``backward'' process that yields, 
when compared to the  forward process of the original dynamics, 
a trajectory-dependent  quantity playing the role of an entropy production
that satisfies a form of the fluctuation theorem.
For systems described by a Langevin dynamics, the  dual backward process
is obtained  easily 
since it is given  by a  Langevin dynamics involving only  additional 
 {\it a priori known} external forces derived from the trial function itself.
   This remarkable property 
offers the possibility  to explore numerically or even 
experimentally  the interesting consequences of the 
associated detailed fluctuation relations, valid 
for systems that spontaneously relax to { nonequilibrium 
steady states}.

The organization of this paper is as follows. In Sec.~\ref{sec:hatanosasa} 
we  review the derivation of the Hatano-Sasa
fluctuation theorem ~\cite{HaSa}.  
After motivating a more general approach we 
provide in Sec.~\ref{sec:hatanosasavariantintegral}
a first derivation of the integral version of our 
fluctuation relation along the same lines of the original Hatano-Sasa 
derivation. In Sec.~\ref{sec:hatanosasavariantdetailed} we give a second, 
more general derivation, 
which yields the detailed version of the theorem 
(containing the integral version as a particular case), and 
in Sec.~\ref{sec:extendeddual} we discuss the physical 
interpretation of the  dual dynamics behind it.
In Sec.~\ref{sec:inequality} we discuss a family of inequalities 
that plays the role of the second law.
In Sec.~\ref{sec:variational} we propose an optimization procedure
for approximating 
steady-state distributions.
As an example, we apply it  to the paradigmatic 
symmetric exclusion process in one dimension. 
In Sec.~\ref{sec:conclusions} we give a conclusion and perspectives.

\section{The Hatano-Sasa relation}
\label{sec:hatanosasa}
Consider a driven system with dynamic variables
${\bf x}$ with  time-dependent
external fields {   $\alpha(t)$}
{  (e.g., shear rate, temperature gradient, etc.)}, with distribution $\rho({\bf x},t)$ 
{  evolving through a generator   $H(\alpha)$:}  
\begin{equation}
 {  \dot \rho({\bf x},t) = -H(\alpha(t)) \; \rho ({\bf x},t)}.
\end{equation}
Let us assume the dynamics admit, for every fixed value of the parameter $\alpha$, a nonequilibrium steady state with distribution
\begin{equation}
{  \rho_{ss}({\bf x};\alpha)=e^{-\phi({\bf {x}};\alpha)}} \;\;\;\;\;\;\;\;  ; \;\;\;\; \;\;\;\; H(\alpha) \rho_{ss}({\bf x};\alpha)=0.
\end{equation}
The Hatano-Sasa~\cite{HaSa} result may be written 
\begin{equation} 
\left\langle e^{-\int_0^{\tau} 
dt \; \frac{\partial \phi({\bf {x}};\alpha)} {\partial \alpha} {\dot{\alpha}}} \right\rangle_{\rho_{ss}(\vx;\alpha_1)} = 1,
\label{equa} \label{wei}
\label{eq:HSv0}
\end{equation}
which implies, by virtue of Jensen's inequality,
\begin{equation}
{ 
\left\langle \int_{\alpha_1}^{\alpha(\tau)} \frac{ \partial \phi({\bf {x}};\alpha)}  {\partial \alpha} d{\alpha} \right\rangle_{\rho_{ss}(\vx;\alpha_1)} \geq 0.
}
\label{inequa}
\end{equation}
The average $\langle \bullet \rangle$  is over all trajectories ${\bf x}(t)$ of  duration $\tau$, 
starting with an initial configuration chosen with the distribution $\rho_{ss}({\bf {x}};\alpha_1)$ with 
$\alpha_1 \equiv \alpha(0)$.
We shall refer to (\ref{equa}) and (\ref{inequa}) as the Hatano-Sasa equality and inequality, respectively.
In the particular case in which the stationary states  $\rho_{ss}(\vx;\alpha)$
are Gibbs states we have
\begin{equation}
\rho_{ss}({\bf x};\alpha) = 
\frac{e^{-\beta E({\bf x},\alpha)}}{e^{-\beta F(\alpha)}} ;
\phi({\bf{x}};\alpha) = \beta [E({\bf x},\alpha) - F(\alpha)]  
\end{equation}
and the Hatano-Sasa equality and inequality become the 
Jarzynski equality and the second law, respectively.

The proof is extremely simple. We start by decomposing the evolution  in a large number $M$ of time steps and compute,
in operator (bra-ket) formalism, the quantity
\begin{eqnarray}
&\langle& -| e^{-\frac{\tau}{M} H(\alpha_M)} 
\frac{ \hat{\rho}_{ss}(\alpha_M)}{ \hat{\rho}_{ss}(\alpha_{M-1})}
... 
\frac{ \hat{\rho}_{ss}(\alpha_3)}{ \hat{\rho}_{ss}(\alpha_2)}
e^{-\frac{\tau}{M} H(\alpha_2)} 
\nonumber \\
& &\frac{ \hat{\rho}_{ss}(\alpha_2)}{ \hat{\rho}_{ss}(\alpha_1)}
 e^{-\frac{\tau}{M} H(\alpha_1)}|\rho_{ss}(\alpha_1)\rangle 
= \langle-|\rho_{ss}(\alpha_M)\rangle=1,
 \label{tele}
\end{eqnarray}
where $e^{-\frac{\tau}{M}H(\alpha_i)}|\rho_{ss}(\alpha_i)\ra=|\rho_{ss}(\alpha_{i+1})\ra$. We denote $\hat{\rho}(\alpha)$ as the operator such that  
$\langle {\bf x}| \hat{\rho}(\alpha) |{\bf x'}  \rangle =  \delta({\bf x}-{\bf x'}) \;\rho({\bf x};\alpha)$
 and  $| \rho(\alpha) \rangle$ as the state such that 
 $\rho({\bf x};\alpha) \equiv \langle {\bf x} |\rho(\alpha)\rangle$.
The state  $|-\rangle$  corresponds to the flat distribution $\langle {\bf x}|-\rangle=1$; 
i.e., the left eigenvector of $H(\alpha)$ has zero eigenvalue.

Now, using that the time step $\tau/M$ is small, we can write
\begin{eqnarray} 
\frac{ {\rho}_{ss}({\bf x};\alpha_{k+1})}{  {\rho}_{ss}({\bf x};\alpha_k)} 
\approx&  e^{-\frac{\tau}{M}  \;  \; \frac{\partial {\phi}({\bf{x}};\alpha_k)}{\partial \alpha}\;  {\dot{\alpha}}}.
\end{eqnarray}
Therefore (\ref{tele}) may be written as Eq. (\ref{equa}), and the result is proven.
The exponential of the term $\left[ -\int dt \; \frac{\partial \phi({\bf x};\alpha)}{\partial \alpha} \;{\dot{\alpha}} \right]$, 
a functional of the trajectory, is thus weighted in (\ref{wei}) with the probability of each dynamical trajectory 
${\bf x}(t)$ such that ${\bf x}(0)$ is sampled from $\rho_{ss}({\bf x},\alpha_1)$. 

In the Hatano-Sasa inequality (\ref{inequa}), 
the equality holds in the quasistationary limit, when the probability distribution may be assumed to be, at each  time $t$, the
  stationary one $\rho_{ss}({\bf x};\alpha)$ corresponding to  the  value  of $\alpha$ at that time:
  \begin{eqnarray}
& &\left\langle
 \int \frac{
\partial \phi({\bf {x}};\alpha)
}
{\partial \alpha}
 d\alpha 
\right\rangle
 = \int  d{\bf{x}} \; d\alpha \; 
 e^{
-\phi({\bf{x}};\alpha)
}
 \frac{
\partial \phi({\bf {x}};\alpha)}{\partial \alpha} \nonumber \\
&=&
\int d {\bf{x}}
[\rho_{ss}({\bf{x}};\alpha_M)-\rho_{ss}({\bf{x}};\alpha_1)] = 0.
\label{HSP1}
\end{eqnarray}
 This result is the generalization of the entropy change 
${S}(\alpha_M)-{S}(\alpha_1)=\langle \int_0^\tau dt\; \dot{\vx} \cdot \nabla\phi\rangle$, 
under reversible transformations, with ${S}(\alpha) \equiv -\int d{\bf x} \rho_{ss}({\bf x};\alpha) 
\ln \rho_{ss}({\bf x};\alpha)$ 
being the generalized Shannon entropy~\cite{HaSa}.  
 
\section{A more general approach}
\label{sec:hatanosasavariant}

The quantity $\phi({\bf x} ; \alpha )= - \ln \rho_{ss}  ({\bf x} ; \alpha )$ plays a role similar to the one of the 
 energy function in a system with detailed balance,   but it may become intractable as soon as we consider a driven system.
The first difficulty  is that  it is, in general, impossible to obtain analytically. This is aggravated by the fact that
in order to use (\ref{equa}) and (\ref{inequa}),  we need to know $\rho_{ss}$ also where it is exponentially small.
Another more serious problem arises from the fact that the function  $\phi({\bf x} ; \alpha )$ may only be small 
in a limited domain and very large everywhere else. An extreme form of this situation  arises in the deterministic limit.
Consider a noisy dynamics  with a (Hoover \cite{Hoover}) thermostat:
\begin{equation}
\left\{
\begin{array}{ll}
\dot q_i &= {p_i},\\ ~\\ \dot p_i &=
 -\frac{\partial { \mathcal H}}{\partial q_i} +
\underbrace{\gamma(t) p_i}_{\text{thermostat}} { 
 - \underbrace{f_i({\bf q})}_{\text{forcing}}     -    \underbrace{\bf \eta_i(t)}_{\text{noise}}}  ,
\end{array}
\right.
\end{equation}
where ${\bf \eta(t)}$ is a Gaussian white noise of variance $\epsilon$. 
Energy is conserved
provided $\gamma(t) = \frac{{\bf (f+{\bm \eta})}\cdot {\bf p}}{{\bf p^2}}$.
 When there is forcing ${\bf f} \neq 0$, the  stationary distribution is  not flat.
Indeed, in the limit of zero noise $\epsilon \to 0$,  
$\rho_{ss}$ has, in fact, fractal support, and $\phi({\bf x};\alpha)$ is infinity almost everywhere on the energy surface.
If we attempt to apply the Hatano-Sasa inequality for a small noise amplitude 
 in a process with varying $\alpha$, 
because the region on the energy shell where $\phi({\bf x};\alpha)$ is small is sparse and strongly dependent on $\alpha$, almost all of the process
takes place in regions in which $\phi({\bf x};\alpha)$ is large: the trajectories are very far from quasistationary, 
and the Hatano-Sasa inequality, though true, becomes useless. 

A similar situation arises when the potential is rapidly oscillating, as in vibrated granular matter, which we may think of as subjected to an oscillating gravity field.
 Here again, the system is always very far from the stationary situation corresponding to any  instantaneous value of the field because it does not have the time to catch up with the oscillating  stationary measure. And yet we still observe  that rapidly vibrated granular matter 
behaves in a manner that resembles motion in contact with a heat bath and would expect some form of second law to apply in that case.

  With the above motivations, we look for  a more flexible approach.
Instead of working with the true stationary distributions $\rho_{ss}({\bf x};\alpha)$, we choose an 
arbitrary family of smooth functions as { reference} states $\rref({\bf x};\alpha)$ and the corresponding 
$\phiref ({\bf x} ;\alpha) \equiv -\ln \rref({\bf x}; \alpha)$. In the following 
we derive an extension of the Hatano-Sasa integral and detailed fluctuation relations, 
 using only these smooth functions.

\subsection{Integral fluctuation theorem}
\label{sec:hatanosasavariantintegral}
In order to obtain a relation, we go through the same steps as  in Sec. \ref{sec:hatanosasa}. 
Starting from the initial distribution 
 $ \phiref ({\bf x} ;\alpha_1) $, we compute, just as in (\ref{tele}),
 \begin{eqnarray}
\langle&-&| e^{-\frac{\tau}{M} H(\alpha_M)} 
\frac{ \hat{\rho}_{\tt ref}(\alpha_M)}{ \tilde {\rho}_{\tt ref}(\alpha_{M-1})}
...
\frac{ \hat{\rho}_{\tt ref}(\alpha_2)}{ \tilde \rho_{\tt ref}(\alpha_1)}
 e^{-\frac{\tau}{M} H(\alpha_1)}|\rref(\alpha_1)\rangle \nonumber \\
&=& \langle-| \rref(\alpha_M)\rangle = 1,
\label{tele1}
\end{eqnarray}
but with $\tilde \rho(\alpha)$ being the operator associated with the state evolved by one time step $e^{-\frac{\tau}{M} H(\alpha)}|\rho(\alpha)\rangle$.
 We can thus write, for large $M$,
\begin{eqnarray} 
\frac{  \tilde{\rho}_{\tt ref}({\bf x};\alpha)}{{\rho}_{\tt ref}({\bf x};\alpha)}  
\approx  e^{ \frac \tau M {\varphi}({\bf x};\alpha)},
\end{eqnarray}
with 
\begin{equation}
\varphi(\vx;\alpha) \equiv -   \frac{1}{\rref(\vx;\alpha)} \; \{ H(\alpha) \; \rref(\vx;\alpha)\}.
\label{eq:varphi}
\end{equation}
Here $H(\alpha)$ acts over the function $\rref(\vx;\alpha)$, so that  it is in fact
 $\varphi(\vx;\alpha) = -\langle \vx| \frac{1}{\hat{\rho}_{\tt ref}(\alpha) } H(\alpha) |\rref(\alpha) \rangle$.
We hence have
\begin{equation}
\frac{ \rref({\bf x};\alpha_{k+1})}{ \tilde \rho_{\tt ref} ({\bf x};\alpha_k)} \approx   e^{-
 \frac{\tau}{M}  \;  \; \frac{\partial \phiref({\bf{x}};\alpha_k)}{\partial \alpha}\;  
{\dot{\alpha}} - \frac \tau M \; {  \varphi(\bf x, \alpha)}},
 \end{equation}
and we obtain a new equality, valid  for all sets $\phiref(\vx;\alpha)$,
\begin{equation}
\left\langle    e^{-
\int dt \; \frac{\partial \phiref ({\bf{x}};\alpha)}{\partial \alpha} \;{\dot{\alpha}} \; \; 
{ - \int dt \; \varphi({\bf x};\alpha)}} \right  
\rangle_{\rref(\vx;\alpha_1)}=1,
\label{eq:HSv1}
\end{equation}
which is the first main result of our paper. Defining 
\begin{equation}
\Yy \equiv \int dt \; \frac{\partial \phiref ({\bf{x}};\alpha)}{\partial \alpha} \;{\dot{\alpha}} \; \; 
{ + \int dt \; \varphi({\bf x};\alpha)} 
\label{eq:entropyproddef}
\end{equation}
it can be simply written as $\langle e^{-\Yy}\rangle=1$.
This integral fluctuation theorem is valid for any protocol $\alpha(t)$ and arbitrary times 
$\tau$, like the Hatano-Sasa equality, to which it reduces if the reference state is chosen as 
$\rref({\bf x};\alpha) = \rho_{ss}({\bf x};\alpha)$, but it holds  for arbitrary smooth functions 
$\rref$ \cite{Sag}. 

As we shall see, this immediately implies an inequality $\langle \Yy \rangle \ge 0$ of the form of the second law.

\subsection{Detailed fluctuation theorem}
\label{sec:hatanosasavariantdetailed}
Just as in the case of  the Hatano-Sasa relation, the result of Eq. (\ref{eq:HSv1}) 
can be alternatively derived as a particular case of a detailed fluctuation 
theorem. We will use here a procedure that generalizes the one used 
for obtaining the detailed fluctuation theorem associated with the 
Hatano-Sasa theorem~\cite{HaSa,chernyak,Esposito,Reinaldo}.


We are looking for a time-reversed form of the dynamics. Let us start by rewriting (\ref{tele1}) as
\begin{eqnarray}
\langle&-&| e^{-\frac{\tau}{M} H(\alpha_M)} 
\frac{ \hat{\rho}_{\tt ref}(\alpha_M)}{ \tilde {\rho}_{\tt ref}(\alpha_{M-1})}
...
\frac{ \hat{\rho}_{\tt ref}(\alpha_2)}{  \tilde {\rho}_{\tt ref}(\alpha_1)}
 e^{-\frac{\tau}{M} H(\alpha_1)}|\rref(\alpha_1)\rangle \nonumber \\
&=& 
\langle {\rho}_{\tt ref}(\alpha_M) | 
\frac{1 }{ \tilde {\rho}_{\tt ref}(\alpha_{M-1})}
...
\frac{ \hat{\rho}_{\tt ref}(\alpha_2)}{  \tilde {\rho}_{\tt ref}(\alpha_1)}
 e^{-\frac{\tau}{M} H(\alpha_1)}{\hat{\rho}}_{\tt ref}(\alpha_1)| - \rangle
\nonumber \\
&=& \langle {\rho}_{\tt ref}(\alpha_M) |  \Pi_{k=1}^{M-1} \left[\frac{ 1}{  \tilde {\rho}_{\tt ref}(\alpha_k)}
 e^{-\frac{\tau}{M} H(\alpha_k)}{\hat{\rho}}_{\tt ref}(\alpha_k)\right] | - \rangle \nonumber.
\end{eqnarray}
We may now take the adjoint in order to reverse time:
\begin{eqnarray} 
  &\langle& - |  \Pi_{k=M-1}^{1} \left[\frac{ 1}{  \tilde {\rho}_{\tt ref}(\alpha_k)}
 e^{-\frac{\tau}{M} H(\alpha_k)}{\hat{\rho}}_{\tt ref}(\alpha_k)\right]^\dag  | {\rho}_{\tt ref}(\alpha_M) \rangle =\nonumber \\
 \langle   - &|&  \Pi_{k=M-1}^{1} \left[\frac{ 1}{ \hat{\rho}_{\tt ref}(\alpha_k)}
 e^{-\frac{\tau}{M} \{H(\alpha_k) + \varphi(\alpha_k)\}}{\hat{\rho}}_{\tt ref}(\alpha_k)\right]^\dag  | {\rho}_{\tt ref}(\alpha_M) \rangle \nonumber \\
 &=&   \langle - |  \Pi_{k=M-1}^{1} \left[
 e^{-\frac{\tau}{M} H^{adj}(\alpha_k) }\right]  | {\rho}_{\tt ref}(\alpha_M) \rangle \nonumber.
\end{eqnarray}
This is a time-reversed dynamics with generator:
\begin{equation}
[H^{adj}(\alpha)]^{\dag} \equiv  \frac{ 1}{   {\hat{\rho}}_{\tt ref}(\alpha)}
\;  \{H(\alpha) + \varphi(\alpha)\} \; {\hat{\rho}}_{\tt ref}(\alpha).
\label{revtim}
\end{equation}
We shall see below that it corresponds, in fact, to a Langevin process with a modified force field [see Eq. (\ref{eq:langevinextendeddual})].

In terms of  the original and the adjoint dynamics, the evolution in time step $\tau/M$ is
\begin{eqnarray}
P(\Bx'|\Bx;\alpha)\equiv \langle \Bx' | e^{- \frac{\tau}{M} {H}_\alpha }  | \Bx \rangle,\\ 
{P^{adj}}(\Bx'|\Bx;\alpha) \equiv \langle \Bx' | e^{-\frac{\tau}{M} {H}^{adj}_\alpha } | \Bx \rangle.
\label{eq:transformeddynamics}
\end{eqnarray}
The construction (\ref{revtim}) tells us that , for each 
trajectory $\Tx\equiv \{ \Bx_1,\Bx_2,...,\Bx_M \}$ with the initial condition chosen with probability
$\rho_{\tt ref}(\Bx_1;\alpha_1)$, there is a time-reversed ($R$) trajectory, with the initial condition 
 chosen with  probability $\rho_{\tt ref}(\Bx_M;\alpha_{M})$, and their respective weights are
\begin{eqnarray}
\Pp[\Tx;\alpha] &=& \prod_{n=1}^{M-1} P(\Bx_{n+1}|\Bx_{n};\alpha_n) \rho_{\tt ref}(\Bx_1;\alpha_{1}), 
\nonumber
\end{eqnarray}
and
\begin{eqnarray}
\left[{\Pp^{{adj} }}[\Tx;\alpha] \right]^R&=& \prod_{n=1}^{M-1} {P^{adj}}(\Bx_{n}|\Bx_{n+1};\alpha_n) \rho_{\tt ref}(\Bx_M;\alpha_{M}). 
 \nonumber 
\end{eqnarray}  
We thus may define a   quantity $ \Xi[\Tx;\alpha]$ associated with each path, having an interpretation analogous to the {\em entropy production},
\begin{equation}
\Xi[\Tx;\alpha] \equiv \ln \frac{\Pp[\Tx,\alpha]}{\left[\Pp^{{adj} }[\Tx,\alpha]\right]^R}.
\label{entropyprod}
\label{eq:v1DFT}
\end{equation}
In the large $M$ limit, it becomes
\begin{equation}
\Xi[\Tx;\alpha] \approx 
\int_{0}^{\tau} dt\;
\left[ {\varphi}(\Bx;\alpha) +
\dot{\alpha} \partial_\alpha \phi_{\tt ref}(\Bx,\alpha)\right].
\label{eq:Xi}
\end{equation}
It is clear from Eq. (\ref{entropyprod}) that, in terms of  $\Xi$, there is a detailed fluctuation theorem:  
\begin{eqnarray}
\langle 
\mathcal{O}[\Tx] e^{-\Xi[\Tx,\alpha]} 
\rangle &=&\langle 
\mathcal{O}[\Tx] e^{-\int_{0}^{\tau} dt\;
\left[ \varphi(\Bx;\alpha) +
\dot{\alpha} \partial_\alpha \phiref(\Bx,\alpha)\right]} 
\rangle \nonumber \\
&=& {\left [{\langle \mathcal{O}[\Tx] \rangle}^{{adj}}\right ]^R},
\label{eq:generalDFT}
\end{eqnarray}
which is valid for an arbitrary functional  $\mathcal{O}[\Tx]$ of the trajectory. 
The averages in (\ref{eq:generalDFT}) are performed with the real forward dynamics in the first term 
and with the time-reversed adjoint dynamics  
of Eq. (\ref{eq:transformeddynamics}) in the second term. 

Equation (\ref{eq:generalDFT}) is a very general result.
It represents a broad family of fluctuation theorems with a trajectory dependent 
``entropy production''  of the form of Eq. (\ref{eq:Xi}), 
completely determined  by the  distributions $\rho_{\tt ref}(\vx;\alpha)$. 

Clearly, choosing $\mathcal{O}=1$ in this equation we get 
the integral fluctuation relation of Eq. (\ref{eq:HSv1}). 
This detailed fluctuation theorem, which can 
be used to derive a variety of Crooks-like relations, is the second 
main result of our paper. 

\subsection{Generalized dual (adjoint) dynamics}
\label{sec:extendeddual}
In order to give a simple physical interpretation of the dual  dynamics let us now 
assume that our system is governed by a Langevin equation,
\begin{equation}
\dot{\Bx}={\bf f}(\Bx;\alpha)+{\bm \xi}(t),
\label{eq:langevin}
\end{equation}
with ${\bf f}(\Bx;\alpha)$ being an arbitrary force (conservative or nonconservative),
and being ${\bm \xi}(t)$ a Gaussian uncorrelated noise at temperature $T$, such that 
$\langle {\bm \xi}(t) \rangle=0$ and $\langle {\xi}_n(t) {\xi}_m(t') \rangle=2T\delta(t-t')\delta_{nm}$. 
This is associated with the Fokker-Planck process:
\begin{equation}
\frac{d\rho}{dt} = \nabla \cdot \left[ \left[ T  \nabla  -  {\bf f}(\vx;\alpha)\right] \rho\right] 
= -H(\alpha) \rho.
 \label{eq:FP}
\end{equation}
Using  Eq.~(\ref{eq:varphi}), $\varphi$ is  given in this case
by 
\begin{equation}
\varphi =   -{\nabla \cdot {\bf f}} -  T{\nabla^2 \phiref} 
+ T \left| {\nabla  \phiref}  \right|^2 +   {\bf f} \cdot {\nabla \phiref}.
\label{eq:varphiFP}
\end{equation}
It is easy to check that, if $e^{-\phiref}$ is the stationary distribution, this expression vanishes.
\\
The expression for a path probability is
\begin{equation}
\Pp[\Tx;\alpha] \sim e^{-\frac{1}{4T}\int_0^\tau dt\; \{[\dot{\Bx}-{\bf f}(\Bx;\alpha)]^2 + 4T \frac{\nabla \cdot {\bf f}}{2}\}}.
\label{eq:weight}
\end{equation}
The last term in the argument of the integral comes from the Stratonovich discretization scheme. 
Then,  using equations (\ref{entropyprod}) and (\ref{eq:entropyproddef}) 
and time reversing in order to obtain the dynamical weight 
(that is, $[\Pp e^{-\Yy}]^R = \left[[\Pp^{{adj}}]^ R\right]^R = \Pp^{{adj}}$), 
we have
\begin{eqnarray}
\label{eq:ddaggerweight}
& &{\Pp^{{adj}}}[{\bf x};\alpha] \sim \\
& &e^{-\frac{1}{4T}\int_{0}^{\tau} dt\;\left[ (\dot{\Bx}+{\bf f}(\Bx;\alpha))^2 + 4T (\varphi(\Bx,\alpha)
- \dot{\alpha} \partial_\alpha \phi_{\tt ref}(\Bx,\alpha) + \frac{\nabla \cdot {\bf f}}{2}) \right]} \sim \nonumber \\ 
& & e^{-\frac{1}{4T}\int_{0}^{\tau} dt\;\left[ (\dot{\Bx}+{\bf f}(\Bx;\alpha))^2 + 4T\varphi(\Bx,\alpha) 
+ 4T\dot{\Bx} \cdot \nabla \phi_{\tt ref}(\Bx,\alpha) + 2T {\nabla \cdot {\bf f}}\right]}, \nonumber
\end{eqnarray}
where in the last step we have dropped all reference to the  boundary  term, 
which is irrelevant for our present purposes.

{\em  Is there a Langevin equation associated with the weight of Eq. (\ref{eq:ddaggerweight})?} 
In order to answer such a question, we follow a procedure analogous to the one used recently in 
Ref.~\cite{Reinaldo} for the standard dual dynamic weight.
Plugging  expression (\ref{eq:varphiFP}) into Eq. (\ref{eq:ddaggerweight}) leads to a simple expression,
\begin{eqnarray}
\label{eq:extendeddualweigth}
{\Pp^{{adj}}}[\Tx;\alpha] \sim e^{-\frac{1}{4T} \int_0^\tau dt\; \left[ (\dot{\Bx}+{\bf f}+ 2T \nabla \phi_{\tt ref})^2 
- 2T\nabla \cdot [{\bf f} +  2T{\nabla \phiref}] \right] }, \nonumber 
\end{eqnarray}
where we can clearly identify the action of the following  Langevin equation (in Stratonovich scheme):
\begin{equation}
{ \dot{\Bx} = -{\bf f}(\Bx;\alpha) - 2T \nabla \phi_{\tt ref}(\Bx;\alpha) + {\bm \xi}(t)}.
\label{eq:langevinextendeddual}
\end{equation}
The dual (adjoint)  dynamics corresponds to a Langevin process  with opposite force 
and an additional external potential $\phi_{\tt ref}(\Bx;\alpha)$ which depends on the choice of $\rho_{\tt ref}$.   

All the results obtained so far reduce to the 
Hatano-Sasa results if we choose $\rref=\rho_{ss}$, in 
which  case, $\varphi=0$,  $\Xi$ becomes the Hatano-Sasa functional 
$\mathcal{Y}_{HS} = \int_{0}^{\tau} dt\;\dot{\alpha} \partial_\alpha \phi(\vx,\alpha)$, 
and the extended dual dynamics becomes the well-known ($\dagger$) 
standard dual dynamics~\cite{bertini,HaSa,Esposito,chernyak}, 
which in terms of transition probabilities reads 
${P^\dagger}(\Bx|\Bx';\alpha) \equiv  
P(\Bx'|\Bx;\alpha) \frac{\rho_{ss}(\Bx,\alpha)}{\rho_{ss}(\Bx';\alpha)}$, 
as can easily be obtained from Eq. ~(\ref{eq:transformeddynamics}). 
The Langevin equation for the usual Hatano-Sasa dual 
dynamics (see for, instance, its derivation in Ref.~\cite{Reinaldo})
coincides with Eq.~(\ref{eq:langevinextendeddual}), 
replacing $\phiref$ by $\phi = -\ln \rho_{ss}$.

Finally, it is worth noting that the extended dual dynamics derived above
has the  advantage over the standard dual dynamics  
that all the forces are known, so that  it might be 
implemented in practice, numerically or even experimentally, by applying 
appropriate external fields. 
It should thus be possible to verify, numerically or 
experimentally, the detailed fluctuation theorem of Eq.~(\ref{eq:v1DFT}) as well as other 
{\it extended} Crooks-like~\cite{Cr}   relations 
that easily follow from Eq.~(\ref{eq:v1DFT}) and concern   
 systems with nonequilibrium steady states. It is with this practical 
motivation that in the Appendix 
we  derive the corresponding extended version of 
the three detailed fluctuation theorems of Esposito and Van der Broeck~\cite{Esposito}. It would 
also be interesting to explore further the implications of the extended dual dynamics,  generalizing the results 
based on  dual dynamics approaches in Refs.~\cite{bertini,HaSa,Esposito,chernyak}).

\subsection{Generalizations of the second law}
\label{sec:inequality}
As we did in Sec. \ref{sec:hatanosasa}
we use the Jensen's inequality in Eq. (\ref{eq:HSv1}) to obtain
\begin{equation}
{ 
\left\langle \int \frac{ \partial \phiref({\bf {x}};\alpha)}  
{\partial \alpha} d{\alpha}  \; \; { + \int dt \; \varphi ({\bf x};\alpha)}
\right\rangle_{\rref(\vx;\alpha_1)}
 \geq 0.
}
\end{equation}
It is worth emphasizing that the time evolution implied in the brackets is the real dynamics 
with initial states drawn from the distribution $\rref(\vx;\alpha_1)$.
The relation is true for arbitrary $\rref({\bf x};\alpha)$; a bad choice only makes the inequality less constraining. 
This is the third main result of our paper 
and the central formula we will exploit for applications.
The function {$\varphi({\bf x};\alpha)$} is a known, well-behaved extensive function of the dynamic variables, which vanishes 
if $\rref({\bf x};\alpha) = \rho_{ss}({\bf x};\alpha)$. For example, for a Langevin
process (\ref{eq:langevin}) it is given by Eq (\ref{eq:varphiFP}).

If at constant $\alpha$ the system is able to converge to a stationary nonequilibrium regime, the inequality 
has to hold for large times such that the initial condition is forgotten. We thus get the stationary-state expectation:
\begin{equation}
 \left\langle \varphi \right\rangle_{ss} = 
 -\left\langle 
 \frac{1}{\rref}{H \;  \rref          }
 \right\rangle_{ss} 
 \ge 0 \;\;\; \forall \rref.
\label{stat}
\end{equation}
This inequality
is already implicit in the work of  Lebowitz and Bergmann~\cite{LeBe}. 
If we define $\dot w_{\tt ref} = -H w_{\tt ref}$ with $w_{\tt ref}(t=0) = \rref$, we can rewrite (\ref{stat}) as
 \begin{eqnarray}
\langle \varphi \rangle_{ss} &=&
-\left[ \frac{d}{dt} \int d{\bf x} \;
\rho_{ss} \ln \left(\frac{\rho_{ss}}{w_{\tt ref}(t)}\right)  \right]_{t=0}
 \nonumber \\
&=& -\left[\frac{d}{dt} D_{KL} (\rho_{ss}\parallel w_{\tt ref}(t))\right]_{t=0}
  \ge 0
 \label{BL}
 \end{eqnarray}
by virtue of the general result $\dot D_{KL}(w_1(t)\parallel w_2(t)) \le 0$, valid for all 
times $t \ge 0$ with $w_1(t)$ and $w_2(t)$ being any two distributions evolving through $H$~\cite{LeBe}. 
The positively defined  Kullback-Leibler distance $D_{KL}$ used above is 
often an  actor in these problems; see \cite{Parrondo,Esposito,Merhav,Lacoste}.

For a purely Hamiltonian system $\dot D_{KL} (w_1(t) \parallel w_2(t)) =  0$
independently of  $w_1$ and $w_2$: 
irreversibility in this case  inescapably requires some form of coarse graining, 
which this method does not provide. 
 Instead, in the case of a Langevin process (\ref{eq:FP}), a short computation~\cite{Risken} gives
\begin{equation}
 \langle \varphi \rangle_{ss} = T \langle |\nabla(\phiref-\phi)|^2\rangle_{ss} \ge 0,
\end{equation}
where $\phi=-\ln \rho_{ss}$. We have easy access to the left-hand side of the above equation numerically {\em or even experimentally} because we know $\phiref$ 
and the dynamics, but we do not have easy access to the right-hand side.

 Let us consider now a system that is perturbed periodically, 
such as the granular system described above. Assume further that the system reaches, 
after a long time, a periodic state. We then have
  \begin{equation}
{ 
\left\langle \oint \frac{ \partial \phiref({\bf {x}};\alpha)}  
{\partial \alpha} d{\alpha}  \; \; { + \oint dt \; \varphi ({\bf x};\alpha)}
\right\rangle
 \geq 0,
}
\end{equation}
 where the time integral is over one cycle in the  regime in which the distribution becomes periodic in time. If we make the further
 simplification that $\rref$ is constant in time, we get
  \begin{equation}
\left\langle  {  \oint dt \; \varphi ({\bf x};\alpha)}
\right\rangle
 \geq 0,
\end{equation}
where the dependence of $\varphi$ on $\alpha$ comes from $H(\alpha)$.

\section{A variational scheme} 
\label{sec:variational}
The preceding section suggests an iterative variational procedure to optimize $\rref$ at fixed $\alpha$:
Propose a change to $\rref({\bf x})$,
compute the new $\varphi= -{\rref^{-1}} H \rref $ (immediate), and
run  $\langle \varphi({\bf x)} \rangle_{ss}$ and accept the change if the result is smaller. The resulting $\varphi$ yields 
directly a second-law-like constraint, which is optimized. 
The optimization procedure we propose might be indeed implemented numerically or even experimentally 
to calculate, for instance, optimal effective interactions from steady-state measurements~\cite{cocco}.

\subsection{An application}
As an illustrative and nontrivial example, we consider the  simple symmetric exclusion process (SSEP),  a one-dimensional 
lattice of $L$ sites that are either occupied by a single particle or
empty. 
A configuration at time $t$ is defined by the vector of occupation numbers
$\mathbf{n}(t)=[n_1(t),...,n_L(t)]$ [$n_i(t)$=0,1].
Each particle in the bulk independently
attempts to jump to an empty site to its  right or to its left.  At the left boundary 
each particle is injected at site 1 at rate
$\alpha$ and removed from site 1 at rate $\gamma$, whereas at the right boundary
particles 
are injected at site $L$ at rate $\delta$ and removed from site $L$ at rate $\beta$. 

\begin{figure}
\centerline{\psfig{file=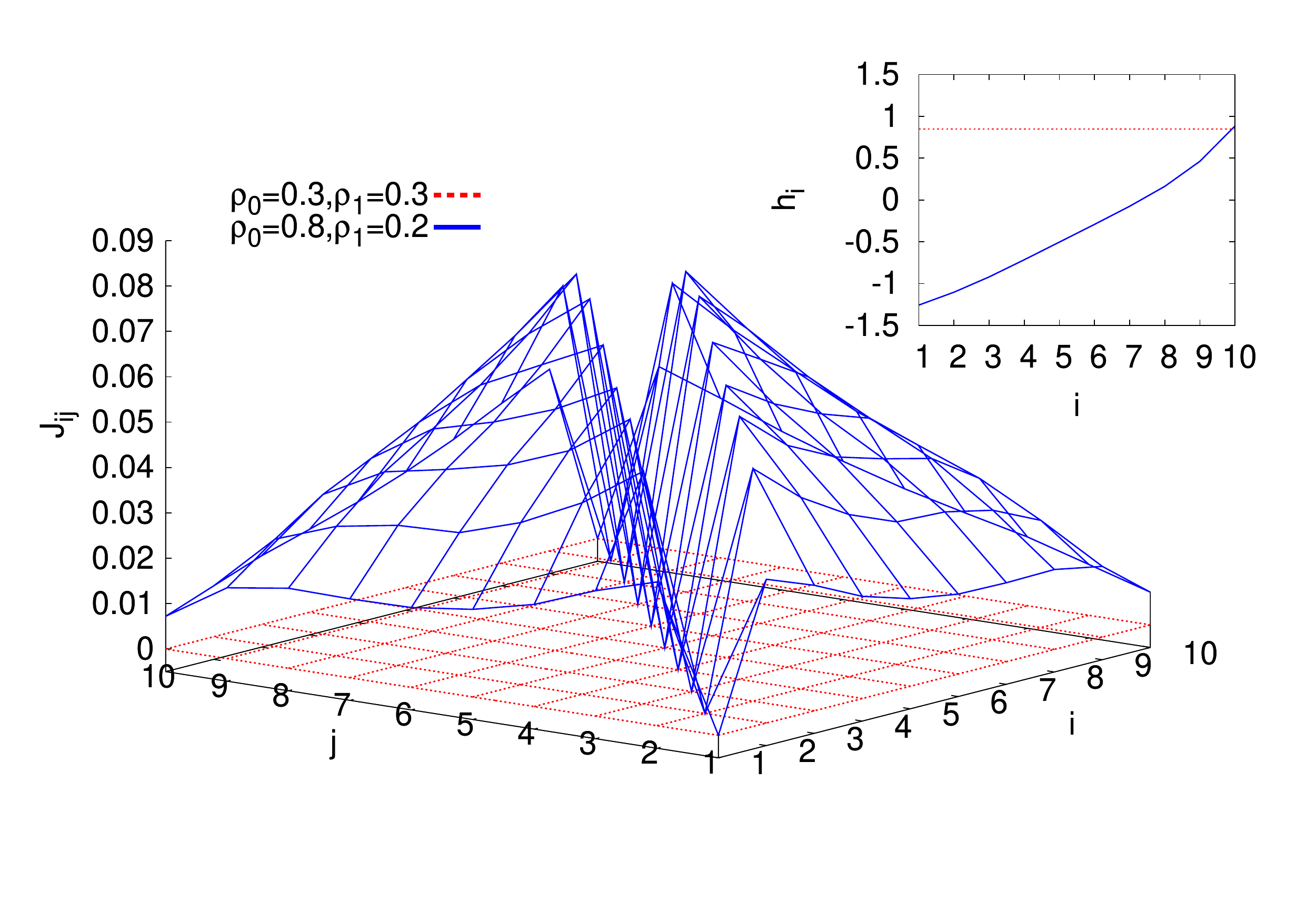,width=9.5cm,clip}}
\caption{(Color online) Optimized $J_{ij}$ for the SSEP model with open boundaries for $\rho_0=\rho_1$ (dotted red line) and for $\rho_0\neq\rho_1$ (solid blue line). 
The inset shows similar results for the optimized $h_i$}
\label{SSEP_L10}
\end{figure}

\begin{figure}[here]
\centerline{\psfig{file=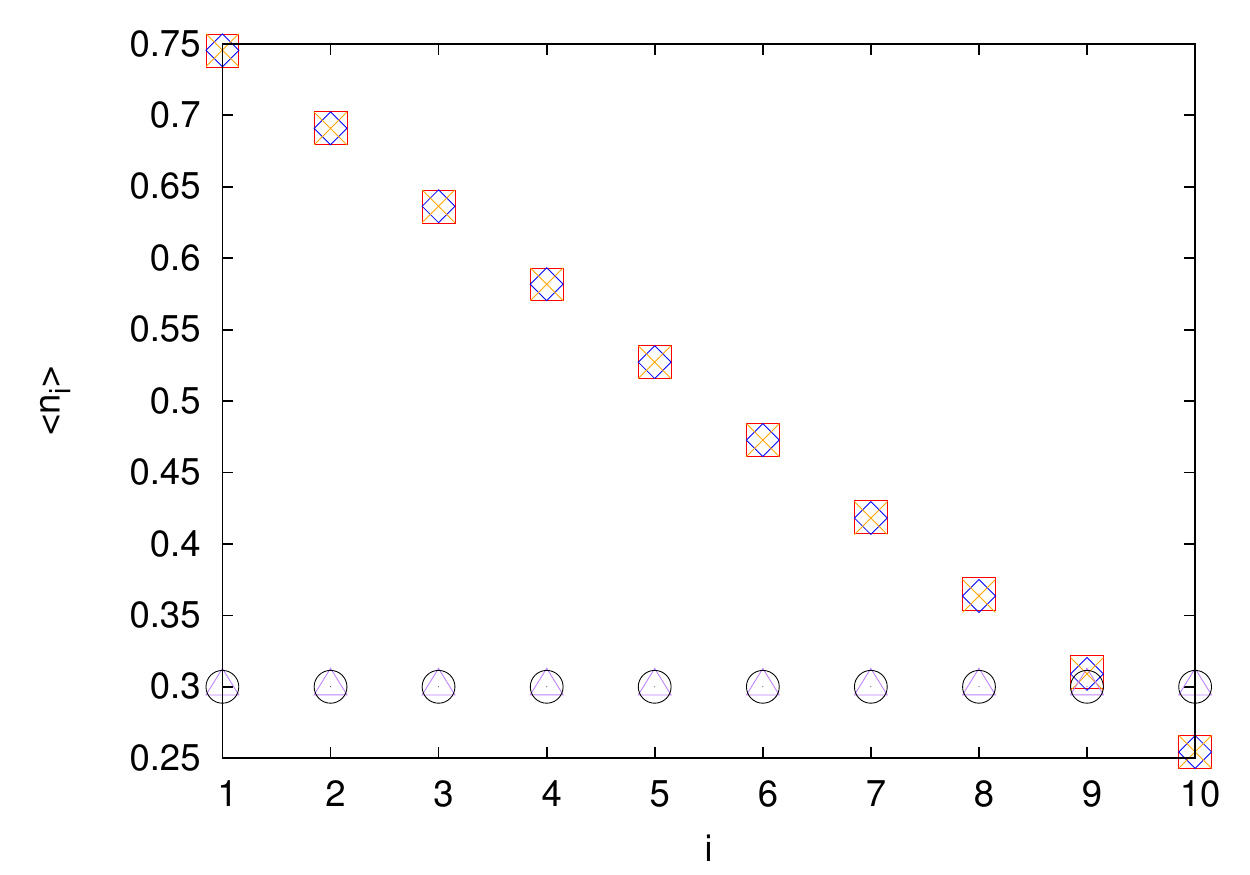,width=9cm,clip}}
\caption{(Color online) Analytical and simulation results for the steady state density profile $\la n_i \ra$. Red squares \textcolor{red}{$\square$} 
and orange crosses \textcolor{orange}{$\times$} correspond to the Monte Carlo procedure for densities $\rho_0=0.8$, $\rho_1=0.2$ using the optimized trial $\rho_{opt}(\vn)$ 
and the local equilibrium measure $\rho_{LE}(\vn)$ respectively, whereas blue diamonds \textcolor{blue}{$\Diamond$} are the analytical 
results, see Eq. (\ref{ssdp}). Purple triangles \textcolor{purple}{$\bigtriangleup$} correspond to the Monte Carlo procedure for densities $\rho_0=\rho_1=0.3$ using the equilibrium 
measure $\rho_{eq}(\vn)$, whereas black circles \textcolor{black}{$\bigcirc$} are the analytical results.}
\label{SSEP_prof1}
\end{figure}

\begin{figure}[here]
\centerline{\psfig{file=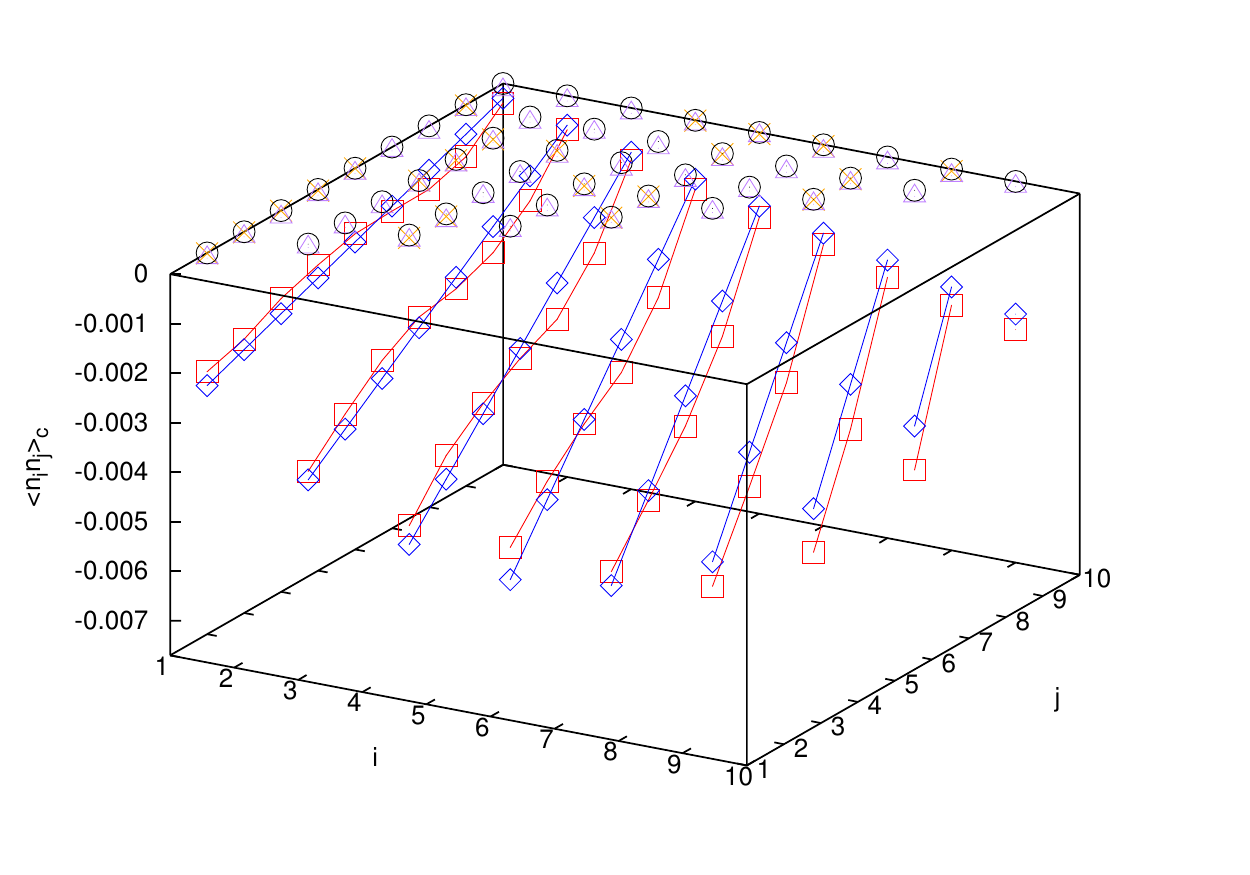,width=10cm,clip}}
\caption{(Color online) Analytical and simulation results for the two-point correlation function $\la n_i n_j \ra_c$ in the steady state.
Red squares \textcolor{red}{$\square$} 
and orange crosses \textcolor{orange}{$\times$} correspond to the Monte Carlo procedure for densities $\rho_0=0.8$, $\rho_1=0.2$ using the optimized trial $\rho_{opt}(\vn)$ 
and the local equilibrium measure $\rho_{LE}(\vn)$ respectively, whereas blue diamonds \textcolor{blue}{$\Diamond$} are the analytical 
results, see Eq. (\ref{sscf}). Purple triangles \textcolor{purple}{$\bigtriangleup$} correspond to the Monte Carlo procedure for densities $\rho_0=\rho_1=0.3$ using the equilibrium 
measure $\rho_{eq}(\vn)$, whereas black circles \textcolor{black}{$\bigcirc$} are the analytical results.}
\label{SSEP_prof2}
\end{figure}

The choice of the rates $\alpha$, $\gamma$, $\delta$ and $\beta$ corresponds to the system being in 
contact with infinite left and right reservoirs at densities $\rho_0=\alpha/(\alpha+\gamma)$ and  $\rho_1=\delta/(\delta+\beta)$, 
respectively \cite{derrida_leb}.
If $\rho_0=\rho_1=\rho$, the system is in equilibrium, and the  distribution is of product form:
$
\rho_{eq}(\mathbf{n})=\prod_{i=1}^{L}\rho^{n_i}(1-\rho)^{1-n_i}=e^{\sum_{i=1}^{L}\mu n_i}/(1+e^{\mu})^L
$,
where $\mu=\ln[\rho/(1-\rho)]$ is the chemical potential.
As soon as $\rho_0 \neq \rho_1$, a current is established, and the problem becomes nontrivial, with long-range correlations. 
The evolution of the probability 
$\rho(\vn)$ of observing a configuration $\vn$ is given by the master equation ($n^+_k=n_k+1$ and $n^-_k=n_k-1$)
\begin{eqnarray}
\begin{array}{lll}
\displaystyle\frac{\partial \rho(\vn)}{\partial t}=\sum_{k=1}^{L-1} [\delta_{n_k,1}\delta_{n_{k+1},0}\rho(...,n_k^{-},n_{k+1}^{+},...) &
\\
+\delta_{n_k,0}\delta_{n_{k+1},1}\rho(...,n_k^{+},n_{k+1}^{-},...) &
\\
-(\delta_{n_k,1}\delta_{n_{k+1},0}+\delta_{n_k,0}\delta_{n_{k+1},1}) \rho(...,n_k,n_{k+1},...)] &
\\
+\alpha \delta_{n_1,1}\rho(n_1^{-},...)+\gamma \delta_{n_1,0}\rho(n_1^{+},...) &
\\
+\delta \delta_{n_L,1}\rho(...,n_L^{-})+ \beta \delta_{n_L,0}\rho(...,n_L^{+}) &
\\
-(\gamma \delta_{n_1,1}+\alpha \delta_{n_1,0}+\beta \delta_{n_L,1}+\delta \delta_{n_L,0})\rho(n_1,...,n_L).
\end{array}
\end{eqnarray}
The full measure on the microscopic configurations in the steady state $\rho_{ss}(\vn)$ may be computed analytically through the so-called 
matrix method \cite{derrida}.
Here we propose an approximate  form  $\phiref({\mathbf{n}})=\sum_{i}h_i
n_i+\sum_{i \neq j}J_{ij}n_in_j$.
Using the master equation, we evaluate $\varphi= \frac{1}{\rref(\mathbf{n})}\frac{\partial \rref(\mathbf{n})}{\partial t}$
[Eq.~(\ref{eq:varphi})] as

\begin{eqnarray}
\varphi&=&\sum_{k=1}^{L-1}
[\delta_{n_k,1}\delta_{n_{k+1},0}e^{h_k-h_{k+1}+\sum_{j \neq k,k+1}2(J_{kj}-J_{k+1,j})
n_j} \nonumber \\
&+& \delta_{n_k,0}\delta_{n_{k+1},1}e^{h_{k+1}-h_k-\sum_{j\neq  k, k+1}2(J_{kj}-J_{k+1,j})
n_j} \nonumber  \\ &-&
(\delta_{n_k,1}\delta_{n_{k+1},0}+\delta_{n_k,0}\delta_{n_{k+1},1})]  \nonumber  \\
&+&
\alpha \delta_{n_1,1}e^{h_1+\sum_{j\neq 1}2J_{1j} n_j}+\gamma
\delta_{n_1,0}e^{-h_1-\sum_{j\neq 1}2J_{1j} n_j}  \nonumber  \\
&+& 
\delta \delta_{n_L,1}e^{h_L+\sum_{j\neq L}2J_{Lj} n_j} +
\beta\delta_{n_L,0}e^{-h_L-\sum_{j\neq L}2J_{Lj} n_j}  \nonumber \\
&-&
(\gamma \delta_{n_1,1}+\alpha \delta_{n_1,0}+\beta \delta_{n_L,1}+\delta
\delta_{n_L,0}).
\end{eqnarray}
We compute the expectation value of this $\varphi$ with the true SSEP dynamics and minimize with respect to
 the $[h_i,J_{ij}]$ using a suitable  algorithm  \cite{Nocedal}.
Clearly, for $\rho_0=\rho_1=\rho$ the system is in equilibrium, and we have for each site 
$h_i=h=-\mu=\ln[(1-\rho)/\rho]$ and $J_{ij}=0$ [see $\rho_{eq}(\vn)$ above]. Unlike the equilibrium case, as soon 
as $\rho_0 \neq \rho_1$,
we obtain nonzero $J_{ij}$ corresponding to the long-range correlations characteristic of the stationary nonequilibrium state; see Fig. \ref{SSEP_L10}. 
These correlations extend over macroscopic distances and reflect the intrinsic nonadditivity of nonequilibrium systems 
\cite{derrida_leb}.
The optimized measure $\rho_{opt}(\vn)=e^{-\phi_{opt}(\vn)}$ obtained with the $[h_i,J_{ij}]$ that minimize the expectation value of 
$\varphi$ is not the exact solution of \cite{derrida}, but we have  checked the quality of the approximation by computing expectation values
with this  measure: this is most easily done with a Monte Carlo procedure with ``energy''
$\phi_{opt}({\bf n})$. To do that one starts from a random initial configuration ${\bf n}(t=0)$ 
and evolves it with a Metropolis algorithm where the probability to go from a configuration ${\bf n}$ to a configuration ${\bf n}'$ in a single 
jump is $W({\bf n} \rightarrow {\bf n}')=\text{min}\left[\displaystyle\frac{\rho_{opt}({\bf n}')}{\rho_{opt}({\bf n})},1\right]$ (note that there are no reservoirs in this calculation).
 The 
configuration ${\bf n}'$ is the same as the configuration ${\bf n}$, except for the randomly chosen node $k$, which changes its value to 
$n_k'=1-n_k$.
We then have
\beq
\displaystyle\frac{\rho_{opt}({\bf n}')}{\rho_{opt}({\bf n})}=\text{exp}\left[(2n_k-1)\left(h_k+
2\sum_{j\neq k}J_{kj}n_j\right)\right].
\eeq
Applying this dynamics we measured the steady-state density profile $\rho_i\equiv \la n_i \ra$ shown in Fig. \ref{SSEP_prof1}, and compared it 
with the analytical result obtained using the exact stationary state measure $\rho_{ss}(\vn)$, which is (see \cite{derrida_leb})
\beq
\la n_i \ra=\displaystyle\frac{\rho_0 (L+\frac{1}{\beta +\delta}-i )+\rho_1(i-1+\frac{1}{\alpha+\gamma})}{L+\frac{1}{\alpha+\gamma}+
\frac{1}{\beta+\delta}-1}.
\label{ssdp}
\eeq
We also 
compared with the result 
obtained assuming local equilibrium considering no reservoirs at the boundaries and a spatially varying chemical potential, which is
adjusted to maintain the same steady-state density profile (\ref{ssdp}). We then have that the local equilibrium measure is 
$\rho_{LE}(\vn)=\prod_{i=1}^Le^{-h_i n_i}/(1+e^{-h_i})$, where $h_i=-\mu_i=\ln[(1-\la n_i \ra)/\la n_i \ra]$, with $\la n_i \ra$ given by 
(\ref{ssdp}). Notice that this local equilibrium measure for $\rho_0\neq \rho_1$ turns into the equilibrium measure by doing $\rho_0=\rho_1$.
In Fig. \ref{SSEP_prof1} we can see that 
there is perfect agreement with the exact analytical results for both $\rho_0=\rho_1$ and $\rho_0\neq\rho_1$ and, in the latter case, for the 
optimized measure and for the local equilibrium measure. 
We also measured the two-point correlation function $\la n_i n_j \ra_c\equiv \la n_in_j\ra-\la n_i\ra \la n_j\ra$, obtaining the results 
shown in Fig. \ref{SSEP_prof2}. Using again the exact measure $\rho_{ss}(\vn)$,one finds that the analytical prediction in the steady 
state for $1\leq i < j \leq L$ is \cite{derrida_leb}
\beq
\la n_in_j\ra_c=\displaystyle\frac{-(\rho_0-\rho_1)^2(\frac{1}{\alpha+\gamma}+i-1)(\frac{1}{\beta +\delta}+L-j)}{(\frac{1}{\alpha+\gamma}+
\frac{1}{\beta+\delta}+L-1)^2(\frac{1}{\alpha+\gamma}+\frac{1}{\beta+\delta}+L-2)}.
\label{sscf}
\eeq
For large $L$, introducing macroscopic coordinates $i=Lx$ and $j=Ly$, this becomes, for $x<y$, $\la n_{Lx}n_{Ly} \ra_c=-x(1-y)(\rho_0-\rho_1)^2/L$. 
As stated in \cite{derrida_leb}, one may think that these weak, but long-range, correlations play no role in the macroscopic limit. However, 
they are responsible for a leading contribution in the variance of a macroscopic quantity such as the number of particles.
\\
As expected, Fig. \ref{SSEP_prof2} shows how the Monte Carlo procedure fits exactly with the analytical results for $\rho_0=\rho_1$ ($\la n_i n_j\ra_c=0$) because we are 
using the equilibrium measure in which no spatial correlations are present. In addition, for $\rho_0\neq\rho_1$ we see how the results obtained 
with the optimized measure are much closer to the exact analytical ones
than those obtained with the local equilibrium assumption. This reflects the fact that with the optimized 
measure we are taking into account at least the two-site long-range correlations, which are not considered in the local equilibrium case. This shows that the physically motivated optimized trial is very good, at least regarding one- and two-point static spatial 
correlation functions. In this sense the out of equilibrium state of the SSEP can be thus fairly approximated by 
simple and intuitive quantities such as the local effective fields and two-site long-range interactions. 
Although more difficult to guess, one might of course add other terms to the trial function to improve the present 
agreement, for instance, higher order interaction terms, but the difficulty of the numerical minimization problem  increases 
very rapidly.
\section{Conclusion}
\label{sec:conclusions}
We have derived an exact relation  for Markovian systems 
that generalizes the Hatano-Sasa relation but  does not rely on  the \textit{a priori} knowledge of the stationary 
probability distribution, but rather on arbitrary trial functions for the stationary distribution.
More generally, we have  derived the detailed version of the fluctuation relation 
by identifying a generalized form of
dual (adjoint) dynamics, generating the  
 backward process that yields
a trajectory-dependent entropy production.
For systems described by  Langevin  dynamics, we have 
showed that the  dual dynamics is  also governed by a simple
Langevin dynamics, which may be expressed  directly in terms  of the trial functions. 
One may also obtain in this context a version of the  three extended detailed fluctuation theorems  of Ref.~\cite{Esposito}.

Our approach leads to an infinite family of inequalities  that generalize the second law, and 
suggests a variational principle for optimizing trial measures, 
in a quantitative and controlled way,  
to approximate  nonequilibrium probability distributions.
The optimization procedure we propose  
might be implemented numerically or even experimentally in order 
to infer nonequilibrium steady-state distributions in 
terms of intuitive physical quantities.
To illustrate this, we have 
implemented this approximating scheme for
the  simple symmetric exclusion 
process in one dimension. 

A particularly interesting case for applying this variational 
approach is to active matter~\cite{unpublished}, where it has been proposed recently 
to represent the  complex energy exchanges in the system by a 
bath with 
equilibrium-like properties (Refs.~\cite{jor,leticia,wolynes}). 
Another interesting and somewhat related system  is that  of  current-driven vortices in superconductors 
with pinning. In this case the complex interplay of driving, 
quenched disorder and vortex-vortex interactions yields a 
variety of nonequilibrium dynamical regimes 
and transitions that  may sometimes be successfully described by 
effective temperatures~\cite{vortices}.
At any rate, the important property of these approximations is that there is a second law-type inequality 
associated with them. 

\section*{ACKNOWLEDGEMENTS}

We acknowledge discussionS with  Cristopher Jarzynski, Shin-Ichi Sasa and Udo Seifert.
A.B.K. acknowledges Universidad de Barcelona, Ministerio de Ciencia e Innovaci\'on (Spain), and 
Generalitat de Catalunya for partial support through the I3 program; C.P.E. acknowledges financial support 
from Junta de Andaluc\'ia (Project No. P07-FQM02725) and the PMMH-ESPCI for its hospitality.

\appendix
\section{Three extended detailed fluctuation theorems}
\label{sec:threeextendeddft}
Using the dual dynamics of  
of Eq. (\ref{eq:langevinextendeddual}), it is straightforward to derive an extension 
for the three fluctuation theorems of Ref.~\cite{Esposito} based, in our case, 
on smooth trial functions $\rref$ rather than
the true steady-state measure $\rss$.

We define three trajectory dependent ``entropy productions'' using time-reversal ($R$) and 
the  dual (adjoint) dynamics:
\begin{eqnarray}
\Ss[\Tx,\alpha] &\equiv& \ln {\Pp[\Tx,\alpha]} - \ln \Pp^R[\Tx,\alpha], \label{eq:Ss} \\
\Yy[\Tx,\alpha] &\equiv& \ln {\Pp[\Tx,\alpha]}-\ln {\left[\Pp^{{adj}}[\Tx,\alpha]\right]^R}, \label{eq:Yy}\\
\Ww[\Tx,\alpha] &\equiv& \ln {\Pp[\Tx,\alpha]}-\ln{\Pp^{adj}[\Tx,\alpha]}, \label{eq:Ww}
\end{eqnarray}
so that by construction they satisfy the detailed fluctuation theorems,
\begin{eqnarray}
\langle 
\mathcal{O}[\Tx] e^{-\Ss[\Tx,\alpha]} 
\rangle 
= {\langle \mathcal{O}[\Tx] \rangle}^{R}, \label{eq:dftSs} \\
\langle 
\mathcal{O}[\Tx] e^{-\Yy[\Tx,\alpha]} 
\rangle 
= \left[{\langle \mathcal{O}[\Tx] \rangle}^{{adj}}\right]^R, \label{eq:dftYy} \\
\langle 
\mathcal{O}[\Tx] e^{-\Ww[\Tx,\alpha]} 
\rangle 
= {\langle \mathcal{O}[\Tx] \rangle}^{{adj}}, \label{eq:dftWw}
\end{eqnarray}
where again $\mathcal{O}$ is an arbitrary functional of the trajectory.

For a system satisfying a Fokker-Planck process (\ref{eq:FP}), it is easy to show 
 using Eqs. (\ref{eq:weight}), (\ref{eq:ddaggerweight}), and (\ref{eq:entropyproddef})
 that
\begin{eqnarray}
\partial_\tau \Ss &=& \beta {\bf f}(\Bx;\alpha)\cdot\dot{\Bx} |_{t=\tau}, \\
\partial_\tau \Yy &=& 
\left[ \dot{\alpha} \partial_\alpha \phiref(\Bx,\alpha)
+ \varphi(\Bx;\alpha) 
\right]|_{t=\tau}, \\
\partial_\tau \Ww &=& 
\left[ \beta {\bf f}(\Bx;\alpha)\cdot\dot{\Bx} - \dot{\alpha} \partial_\alpha \phiref(\Bx,\alpha)
+ \varphi(\Bx;\alpha) \right]|_{t=\tau},
\nonumber \\
\end{eqnarray}
with $\beta \equiv 1/T$. Therefore the total trajectory-dependent entropy production~\cite{Ku} may be written as
\begin{equation}
 \partial_\tau \Ss = \partial_\tau \Yy + \partial_\tau \Ww - 2\varphi.
\end{equation}

The corresponding integral fluctuation theorems for $\Ss$, $\Yy$, and $\Ww$, particular 
cases of Eqs. (\ref{eq:dftSs}), (\ref{eq:dftYy}), and (\ref{eq:dftWw}), respectively, read
\begin{eqnarray}
\langle e^{-\Ss} \rangle =  \langle e^{-\Yy} \rangle = \langle e^{-\Ww} \rangle = 1.
\label{eq:integralextendedfts}
\end{eqnarray}
The first fluctuation theorem in Eq.(\ref{eq:integralextendedfts}) is the well-known integral 
fluctuation theorem for the total entropy production or  
Jarzynski relation, here generalized to arbitrary initial conditions and dynamics without 
detailed balance. The second one 
is the Hatano-Sasa extension discussed in detail in Sec.~\ref{sec:hatanosasavariant}, and the third 
one can be considered as an extension of the Speck and Seifert integral fluctuation theorem~\cite{SpSe}. 
We note, however, that 
the interpretation of $\Yy$ and $\Ww$ as the nonadiabatic $\Ss_{na}$ and adiabatic $\Ss_{a}$ entropy production 
contributions, respectively, is lost due to the presence of $\varphi$, which does not vanish in 
the adiabatic limit if $\rref \neq \rss$. 

When  $\rref = \rss$, $\phiref=\phi$, and $\varphi=0$, and  the three extended fluctuation 
theorems  reduce to the three fluctuation relations described in Ref.~\cite{Esposito}. 
In this case, the dual dynamics becomes exactly the standard 
dual or adjoint ($\dagger$) dynamics, and up to boundary  terms we can identify
$\Yy = \Ss_{na}$, $\Ww = \Ss_{a}$.  In addition, we note that 
Eq.~(\ref{eq:langevinextendeddual}) reduces to the standard dual dynamics in its Langevin form
by the simple replacement $\phiref \to \phi$. The  dual dynamics given here is therefore 
the natural and probably simplest generalization of the standard dual dynamics that does not rely 
on the knowledge of $\rss$.


\begin{thebibliography}{99}
\bibitem{foot1}The Hatano-Sasa, together with 
Seifert~\cite{Seifert} 
and Speck-Seifert relations~\cite{SpSe} 
were very 
recently unified in three detailed fluctuation theorems 
through particular two-fold decompositions of the total 
entropy production, either using single~\cite{Esposito} or 
joint~\cite{Reinaldo} probability distributions for the entropy 
production contributions.
\end{thebibliography}
\end{document}